\pdfoutput=1
\documentclass[aps,twocolumn, floatfix, superscriptaddress]{revtex4}
\usepackage{graphicx}
\DeclareGraphicsExtensions{.pdf}
\DeclareMathAlphabet\mathbfcal{OMS}{cmsy}{b}{n}
\usepackage{amsmath,amssymb,bbold,bm,color}
\usepackage{float}
\usepackage{epstopdf}
\usepackage{tabularx}
\usepackage{braket}

\begin{document}

\title{Strain-induced pseudomagnetic field and quantum oscillations in kagome crystals}

\author{Tianyu Liu}
\affiliation{Max-Planck-Institut f\"ur Physik komplexer Systeme, 01187 Dresden, Germany}

\begin{abstract} 
A kagome lattice is composed of corner-sharing triangles arranged on a honeycomb lattice such that each honeycomb bond hosts a kagome site while each kagome triangle encloses a honeycomb site. Such close relation implies that the two lattices share common features. We predict here that a kagome crystal, similar to the honeycomb lattice graphene, reacts to elastic strain in a unique way that the bulk electronic states in the vicinity of Dirac points are reorganized by the strain-induced pseudomagnetic field into flat Landau levels, while the degenerate edge states in the undeformed crystal become separated in the energy dimension. When the strain is tuned continuously, the resulting scanning pseudomagnetic field gives rise to quantum oscillations in both density of states (DOS) and electric conductivity.
\end{abstract}
\date{\today}

\maketitle
\section{Introduction}
A kagome lattice is a hexagonal Bravais lattice with a 3-site basis. The unusual lattice site arrangement renders the kagome lattice an ideal platform to study geometric frustration \cite{lacroix2011} and the resulting exotic quantum states of matter known as quantum spin liquids \cite{lee2007, balents2010, yan2011, jiang2012, han2012, norman2016, liao2017, zhou2017, lauchli2019}. Also for structural reasons, the wave function associated with a hexagonal ring in the kagome lattice becomes completely localized due to the destructive interference of the wave functions of the corner sites \cite{bergman2008,yang2014}, resulting in highly degenerate dispersionless bands \cite{mielke1991, mielke1992} stable against disorder \cite{bilitewski2018}. The existence of these flat bands has recently been verified using the scanning tunneling microscopy (STM) in layered silicene \cite{li2018} and Co$_3$Sn$_2$S$_2$ \cite{yin2019}. In the presence of spin-orbit interaction and time reversal symmetry breaking, the band degeneracy is lifted and the flat bands acquire a nonzero Chern number, giving rise to the fractional quantum Hall effect (FQHE) when partially filled \cite{tang2011}. The spin-orbit interaction can also gap out the two dispersive bands above the flat band, producing the quantum anomalous Hall effect (QAHE) \cite{zhang2011, xu2015} or the quantum spin Hall effect (QSHE) \cite{guo2009, wang2010}.

In the absence of the spin-orbit interaction and next nearest neighbor hoppings, the two dispersive bands linearly touch at the corners of the Brillouin zone. Such band crossings have been theoretically acknowledged for a long time but experimentally observed only recently in FeSn \cite{kang2019}. Materials with linear energy band crossings react to elastic strain in a unique way by generating in the vicinity of the crossings a chiral gauge field, which is first proposed in graphene \cite{guinea2010,  vozmediano2010, levy2010} and later generalised to Dirac and Weyl semimetals \cite{cortijo2015, sumiyoshi2016, pikulin2016, grushin2016, cortijo2016, arjona2017, liu2017a}, Weyl and $d$-wave superconductors \cite{liu2017b, matsushita2018, kobayashi2018, massarelli2017, nica2018}, and various bosonic Dirac materials \cite{rechtsman2013, ferreiros2018, liu2019}. However, whether similar strain-induced gauge field occurs in kagome lattice has not yet been systematically investigated before, presumably because the additional quadratic band touching of the dispersive bands and the flat band \cite{du2017, ren2018} produces an extra degree of freedom obscuring the Dirac physics derived from the dispersive bands.

In this paper, by projecting to a subspace associated with the linear band crossings, we are able to get rid of the extra degrees of freedom and extract the Dirac physics. Through a combination of analytical calculation and numerical simulation, we elaborate that a properly engineered strain can induce a uniform pseudomagnetic field, which generates flat Landau levels and quantum oscillations. To support these findings, we organize this paper as follows. In Sec.~\ref{sec2}, we study the band structure of a kagome crystal with only nearest neighbor terms and analytically extract the Dirac physics from the sublattice space Hamiltonian. In Sec.~\ref{sec3}, we investigate the response of a kagome crystal to elastic strain with our focus on the bulk Dirac cones as well as the edge states residing on a pair of sawtooth boundaries. In Sec.~\ref{sec4}, we demonstrate the quantum oscillations of the density of states (DOS) and the longitudinal electric conductivity in the kagome crystal. Sec.~\ref{sec5} concludes the paper and discusses the experimental implementation of the strain-induced quantum oscillations.

\section{Model}
\label{sec2}
\begin{figure}[htb]
\includegraphics[width = 8.0cm]{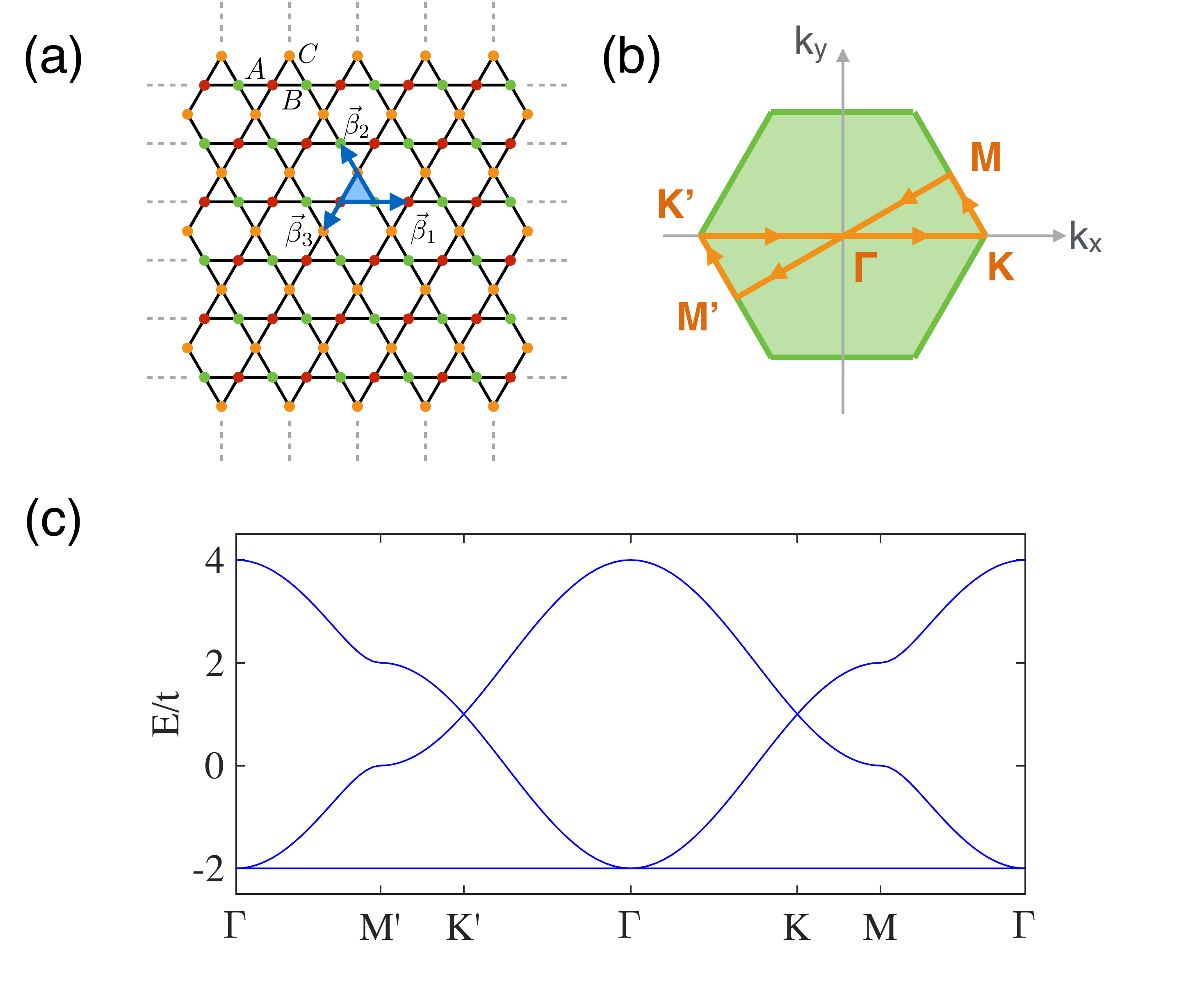}
\caption{(a) Schematic plot of a periodic kagome lattice which is a hexagonal Bravais lattice with a 3-site basis illustrated by the shaded triangle. The associated primitive vectors are $\bm \beta_1 = 2a \hat x$, $\bm \beta_2=-a\hat x + \sqrt{3}a \hat y$, and $\bm \beta_3 = -a\hat x - \sqrt{3} a\hat y$, with $a$ being the lattice constant. (b) The first Brillouin zone of the kagome lattice with high-symmetry points labelled. (c) Band structure of a kagome crystal plotted along the path in (b). There are two linear band crossings at the corners $(K/K')$ of the Brillouin zone and one quadratic band crossing in the center $(\Gamma)$ of the Brillouin zone.} \label{Fig1}
\end{figure}

We consider a toy model of a periodic kagome crystal with only nearest neighbor terms. The unit cell of the lattice contains three sites as illustrated by the shaded triangle in Fig.~\ref{Fig1}(a). The tight-binding Hamiltonian then reads
\begin{equation} \label{H_periodic}
\begin{split}
H=\sum_{\bm r} (t_1 b_{\bm r}^\dagger a_{\bm r} &+ t_1 a_{\bm r + \bm \beta_1}^\dagger b_{\bm r} + t_2 c_{\bm r}^\dagger b_{\bm r}  + t_2 b_{\bm r + \bm \beta_2}^\dagger c_{\bm r} 
\\
&+ t_3 a_{\bm r}^\dagger c_{\bm r} + t_3 c_{\bm r + \bm \beta_3}^\dagger a_{\bm r}) + \text{H.c.},
\end{split}
\end{equation}
where the summation of $\bm r$ runs over all unit cells and $t_{i=1,2,3}$ is the hopping parameter between the nearest neighbors along the direction of the primitive vector $\bm \beta_{i=1,2,3}$. Apply Fourier transform
\begin{equation}
\begin{pmatrix}
a_{\bm r} \\ b_{\bm r} \\c_{\bm r}
\end{pmatrix}
= \frac{1}{\sqrt{N_{\text{uc}}}} \sum_{\bm k} e^{i \bm k \cdot \bm r}
\begin{pmatrix}
a_{\bm k} \\ b_{\bm k} \\c_{\bm k}
\end{pmatrix},
\end{equation}
where $N_{\text{uc}}$ is the number of unit cells. We can then rewrite the Hamiltonian (Eq.~\ref{H_periodic}) in the sublattice basis $\psi_{\bm k} = (a_{\bm k}, b_{\bm k}, c_{\bm k})^T$ as $H=\sum_{\bm k} \psi_{\bm k}^\dagger \mathcal{H}_{\bm k} \psi_{\bm k}$, where the first quantized Hamiltonian matrix reads
\begin{equation} \label{H_Bloch}
\mathcal{H}_{\bm k}=
\begin{pmatrix}
0 && t_1+t_1 e^{-i \bm k \cdot \bm \beta_1} && t_3 + t_3 e^{i \bm k \cdot \bm \beta_3}
\\
 && 0 && t_2+t_2 e^{-i \bm k \cdot \bm \beta_2}
\\
 &&  && 0
\end{pmatrix}.
\end{equation}
In the absence of anisotropy, we have $t_{i=1,2,3}=t$, in which case, $\mathcal{H}_{\bm k}$ produces three bands
\begin{equation}
\begin{split}
\epsilon_{\bm k, 0}&=-2t,
\\
\epsilon_{\bm k, 1}&=t - t \sqrt{3+2\lambda_{\bm k}},
\\
\epsilon_{\bm k, 2}&=t + t \sqrt{3+2\lambda_{\bm k}},
\end{split}
\end{equation}
where $\lambda_{\bm k} = \sum_i \cos (\bm k \cdot \bm \beta_i)$. The flat band $\epsilon_{\bm k, 0}$ results from the destructive interference in the hexagonal rings \cite{yang2014, bergman2008}, while the two dispersive bands $\epsilon_{\bm k,1}$ and $\epsilon_{\bm k,2}$ are similar to the bands of graphene \cite{neto2009, sarma2011}, whose Bravais lattice is also hexagonal but with a 2-site basis. Band $\epsilon_{\bm k,1}$ and band $\epsilon_{\bm k,2}$ cross at the corners of the Brillouin zone $\bm K_\eta = (\eta \tfrac{2\pi}{3a},0)$ at energy $\epsilon_{\bm K_\eta,i=1,2}=t$. To get insights on the band crossings, we study the Hamiltonian in the vicinity of the Brillouin zone corners by projecting $\mathcal{H}_{\bm K_\eta + \bm q}$ onto the space spanned by the eigenvectors $\ket {\phi_{\bm K_\eta,i=1,2}}$ of $\mathcal{H}_{\bm K_\eta}$ associated with eigenenergies $\epsilon_{\bm K_\eta,i=1,2}=t$. Explicitly, 
\begin{multline} \label{H_Dirac} 
\langle \mathcal{H}_{\bm K_\eta + \bm q} \rangle_\phi =
\\
\begin{pmatrix}
\bra{\phi_{\bm K_\eta,1}} \mathcal{H}_{\bm K_\eta + \bm q} \ket{\phi_{\bm K_\eta,1}} && \bra{\phi_{\bm K_\eta,1}} \mathcal{H}_{\bm K_\eta + \bm q} \ket{\phi_{\bm K_\eta,2}}
\\
\bra{\phi_{\bm K_\eta,2}} \mathcal{H}_{\bm K_\eta + \bm q} \ket{\phi_{\bm K_\eta,1}} && \bra{\phi_{\bm K_\eta,2}} \mathcal{H}_{\bm K_\eta + \bm q} \ket{\phi_{\bm K_\eta,2}}
\end{pmatrix}
\\
\approx t\sigma^0+ \sqrt{3} \eta a t q_x \sigma^x + \sqrt{3} a t q_y \sigma^y,
\end{multline}
where we have used $\ket {\phi_{\bm K_\eta,1}}=\tfrac{1}{\sqrt{3}}(1,1,1)^T$ and $\ket {\phi_{\bm K_\eta,2}}=\tfrac{1}{\sqrt{3}}(\tfrac{1}{2}+\tfrac{\sqrt{3}}{2}i\eta, \tfrac{1}{2}-\tfrac{\sqrt{3}}{2}i\eta,-1)^T$. And $\sigma^{x,y}$ and $\sigma^0$ are the Pauli matrices and the unity matrix in the space $\{\ket{\phi_{\bm K_\eta,1}}, \ket{\phi_{\bm K_\eta,2}} \}$, in which $\langle \mathcal{H}_{\bm K_\eta + \bm q} \rangle_\phi$ is a standard Dirac Hamiltonian with velocity parameters $(v_x^\eta, v_y^\eta)=(\sqrt{3}\eta at /\hbar, \sqrt{3} at/\hbar)$.

\begin{figure*}
\includegraphics[width = 17.0cm]{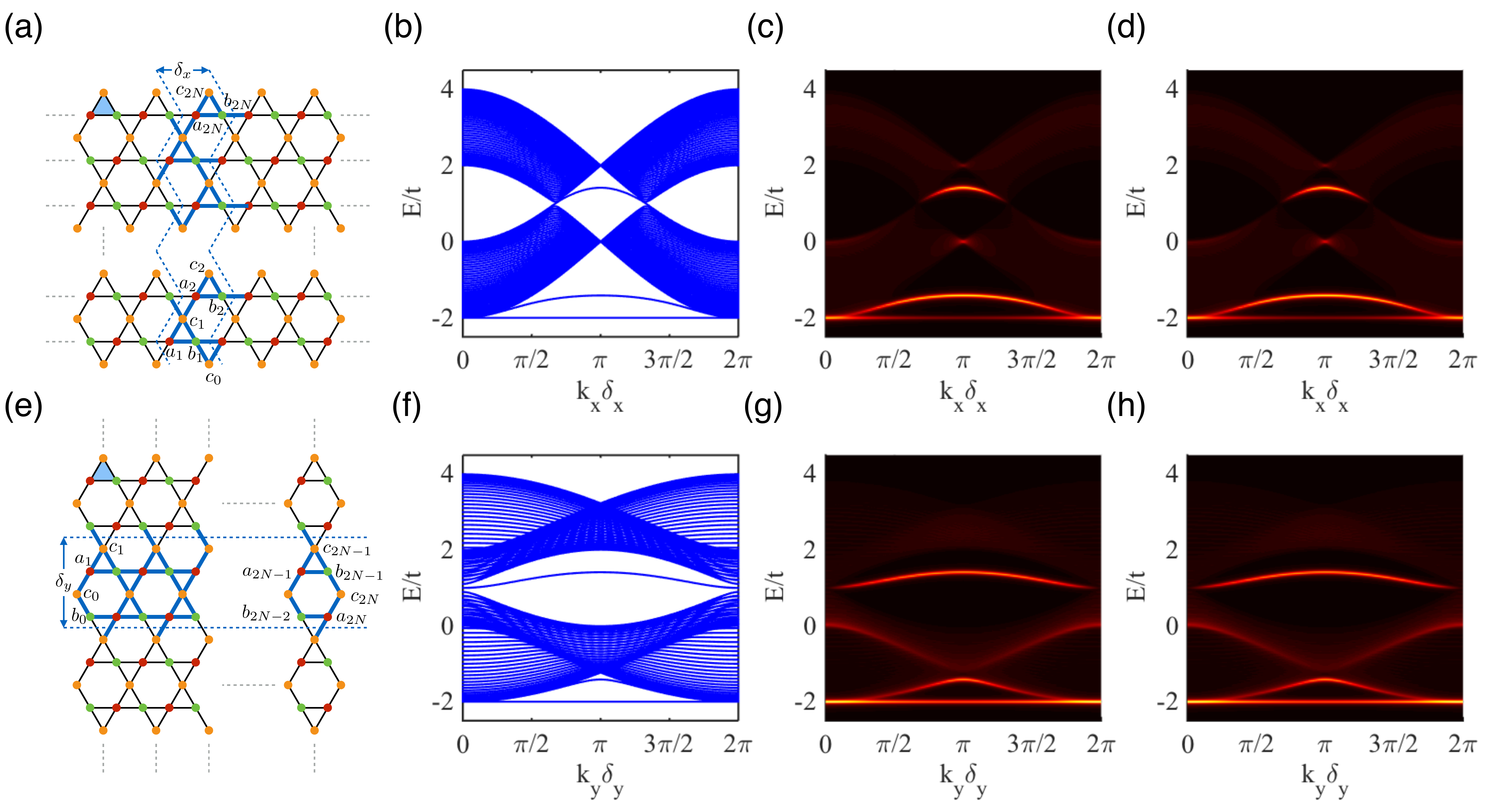}
\caption{Band structure of kagome nanoribbons. (a) A kagome nanoribbon with a pair of sawtooth edges along the $x$ direction. Each unit cell (between the two dashed zigzag lines) contains $2N=50$ shaded triangles and an extra $c$-site $c_0$ on the lower edge. (b) The band structure of the sawtooth kagome nanoribbon in (a). Each sawtooth edge hosts two edge states. One of them is located between the two dispersive bands, connecting the Dirac points, while the other is located between the flat band and the lower dispersive band. (c) The spectral function (Eq.~\ref{A_sawtooth}) of the upper sawtooth edge. (d) The spectral function (Eq.~\ref{A_sawtooth}) of the lower sawtooth edge. (e) A kagome nanoribbon with a pair of zigzag edges along the $y$ direction. Each unit cell (between the two dashed lines) contains $2N-1=49$ shaded triangles. On the left edge, there are an extra $b$-site $b_0$ and an extra $c$-site $c_0$. And on the right edge, there are an extra $a$-site $a_{2N}$ and an extra $c$-site $c_{2N}$. (f) The band structure of the zigzag kagome nanoribbon in (e). Each zigzag edge hosts two edge states. One of them is located between the two dispersive bands, while the other is located between the flat band and the lower dispersive band. (g) The spectral function (Eq.~\ref{A_zigzag}) of the left zigzag edge. (h) The spectral function (Eq.~\ref{A_zigzag}) of the right zigzag edge. It is worth noting that the flat bands in all spectral function plots are bulk states rather than edge states. The appearance of the flat bands can be attributed to the fact that both the sawtooth edges and the zigzag edges contain sites from the hexagonal rings where the flat states are trapped.} \label{Fig2}
\end{figure*}

We have extracted the Dirac physics from the bulk of the sublattice Hamiltonian (Eq.~\ref{H_periodic}). We now briefly discuss the kagome crystals with open boundary conditions. For a nanoribbon with a pair of sawtooth boundaries [Fig.~\ref{Fig2}(a)], we observe that a doubly degenerate arc state connects the two Dirac points while the other doubly degenerate arc state emerges from and terminates at the quadratic band touching arising from the flat band and the lower dispersive band [Fig~\ref{Fig2}(b)]. The edge origination of these arc states can be confirmed by calculating the spectral function on each sawtooth edge
\begin{equation} \label{A_sawtooth}
A_s(E,k_x)=-\frac{1}{\pi} \sum_{y\in \text{edge}} \lim_{\delta \rightarrow 0} \Im [E+ i \delta -\mathcal{H}_{yy'}(k_x)]^{-1}_{y=y'}.
\end{equation}
The spectral function associated with the upper edge is plotted in Fig.~\ref{Fig2}(c), while the spectral function associated with the lower edge is identical as illustrated in Fig.~\ref{Fig2}(d). For a nanoribbon with a pair of zigzag edges [Fig.~\ref{Fig2}(e)], we also find two pairs of doubly degenerate edge states connecting Dirac points and the quadratic band touchings [Fig.~\ref{Fig2}(f)], respectively. The edge origination is confirmed by calculating the spectral function on each zigzag edge
\begin{equation} \label{A_zigzag}
A_z(E,k_y)=-\frac{1}{\pi} \sum_{x\in \text{edge}} \lim_{\delta \rightarrow 0} \Im [E+ i \delta -\mathcal{H}_{xx'}(k_y)]^{-1}_{x=x'}.
\end{equation}
which is plotted in Fig.~\ref{Fig2}(g) [Fig.~\ref{Fig2}(h)] for the left (right) edge.

\section{Strain-induced pseudomagnetic fields}
\label{sec3}
In Sec.~\ref{sec2}, we have studied the band structure of kagome crystals. The linear energy band crossings at Brillouin zone corners indicate that kagome  crystals belong to the family of Dirac matter. One of the most important features of Dirac matter is that the elastic strain is equivalent to a gauge field inducing Landau quantization \cite{guinea2010,  vozmediano2010, levy2010, cortijo2015, sumiyoshi2016, pikulin2016, grushin2016, cortijo2016, arjona2017, liu2017a, liu2017b, matsushita2018, kobayashi2018, massarelli2017, nica2018, rechtsman2013, ferreiros2018, liu2019}. In this section, we will study the reaction of kagome crystals to a properly engineered strain.

The most important effect of the strain is that it alters the position of each lattice site, resulting in spatial modulation of the overlap of electron clouds from neighboring sites \cite{shapourian2015}. Explicitly, the hopping $t_{\bm R', \bm R}$ between site $\bm R'$ and $\bm R$ will be varied by an amount of $\delta t_{\bm R', \bm R}= t_{\bm R' + \bm u(\bm R'), \bm R + \bm u(\bm R)} - t_{\bm R', \bm R}$. To the linear order, it reads
\begin{multline}\label{t_general}
\delta t_{\bm R', \bm R} \approx (\bm R' - \bm R)\cdot \nabla_{\bm r} \bm u(\bm r) |_{\bm R} \cdot \nabla_{\bm r' - \bm r} t_{\bm r', \bm r}|_{\bm R' - \bm R}
\\
= - \frac{g t_{\bm R', \bm R}}{|\bm R' - \bm R|^2} (\bm R' - \bm R) \cdot \nabla_{\bm r} \bm u(\bm r) |_{\bm R}\cdot (\bm R' - \bm R),
\end{multline}
where $\bm u(\bm R)$ is the displacement of the lattice site located at $\bm R$ and we have adopted exponentially decaying overlap integrals $t_{\bm r', \bm r}=t_{\bm R', \bm R} \exp [-g(|\bm r'-\bm r|-|\bm R'-\bm R|)]/|\bm R' -\bm R|$, in which $g$ is the Gr\"unisen parameter of order unity \cite{vozmediano2010}. Without loss of generality, we use $g=1$ in the following. Consequently, in the presence of strain, the overlap integrals in Eq.~\ref{H_periodic} are modulated as $t_i \rightarrow t_i + \delta t_i$ with
\begin{equation} \label{delta_t}
\begin{split}
\delta t_1&=-tu_{xx},
\\
\delta t_2 &=-t (\tfrac{1}{4}u_{xx} + \tfrac{3}{4} u_{yy} - \tfrac{\sqrt{3}}{2} u_{xy}),
\\
\delta t_3 &=-t (\tfrac{1}{4}u_{xx} + \tfrac{3}{4} u_{yy} + \tfrac{\sqrt{3}}{2} u_{xy}),
\end{split}
\end{equation}
where $u_{ij}=\frac{1}{2}(\partial_i u_j+\partial_j u_i)$ is the symmetrized strain tensor whose value should be taken at lattice site $\bm R$, from which an electron hops to the neighboring sites $\bm R'=\bm R+ \bm \beta_i/2$. Under strain, the first quantized Hamiltonian (Eq.~\ref{H_Bloch}) acquires an extra term
\begin{equation} \label{dH}
\delta \mathcal{H}_{\bm k}=
\begin{pmatrix}
0 && \delta t_1+ \delta t_1 e^{-i\bm k \cdot \bm \beta_1} &&
\delta t_3 + \delta t_3 e^{i\bm k \cdot \bm \beta_3}
\\
&& 0 &&\delta t_2 + \delta t_2e^{-i\bm k \cdot \bm \beta_2}
\\
&& && 0
\end{pmatrix}.
\end{equation}
To figure out how this term can affect the Dirac cones, we project $\mathcal{H}_{\bm K_\eta + \bm q} + \delta \mathcal{H}_{\bm K_\eta + \bm q}$ onto the space spanned by $\ket{\phi_{\bm K_\eta,i=1,2}}$ and obtain
\begin{multline} \label{H_strain}
\langle \mathcal{H}_{\bm K_\eta +\bm q} +  \delta \mathcal{H}_{\bm K_\eta +\bm q} \rangle_\phi \approx 
\sqrt{3} \eta at \Big( q_x + \frac{e}{\hbar} \mathcal{A}_x \Big) \sigma^x 
\\
+ \sqrt{3} at \Big( q_y + \frac{e}{\hbar} \mathcal{A}_y \Big) \sigma^y +  (t+\mathcal U) \sigma^0 .
\end{multline}
Consequently, the strain alters the Dirac Hamiltonian in two ways. Firstly, the onsite energy acquires an extra term $\mathcal{U}=-\frac{1}{2}t(u_{xx}+u_{yy})$, which may be interpreted as a strain-induced elastic scalar potential whose gradient corresponds to a strain-induced pseudoelectric field $\mathbfcal{E}=-\frac{1}{e}\nabla \mathcal{U} = \frac{t}{2e} \nabla(u_{xx}+u_{yy})$. Secondly, compared to the Dirac Hamiltonian (Eq.~\ref{H_Dirac}), the Dirac points of this projected Hamiltonian are shifted in the momentum space through a standard Peierls substitution $\bm q \rightarrow \bm q+\frac{e}{\hbar} \mathbfcal{A}$. And the strain-induced elastic vector potential can be read off as
\begin{equation} \label{pseudo_A}
\mathbfcal{A}= \eta \frac{\hbar}{2\sqrt{3}ea} (u_{yy}-u_{xx}, 2u_{xy}).
\end{equation}
In contrast to the ordinary magnetic vector potential $\bm A$ that shift all momenta $\bm k$ in the Brillouin zone through Peierls substitution $\bm k \rightarrow \bm k + \frac{e}{\hbar} \bm A$, the elastic vector potential $\mathbfcal{A}$ only shifts the momenta $\bm q$ in the vicinity of Dirac points. Moreover, $\mathbfcal{A}$ takes opposite signs at different Dirac points due to the valley index $\eta$ in Eq.~\ref{pseudo_A}. But similar to the ordinary magnetic vector potential, $\mathbfcal{A}$ can Landau-quantize electron bands if $\nabla \times\mathbfcal{A} \neq 0$. This requires the strain tensors $u_{ij}$ to be space-dependent which seems to contradict with the aforementioned derivation assuming constant strain. However, we argue that even when $u_{ij}$ is spatially varying, the strain effect can still be treated as an elastic gauge field, as long as it varies slowly on the lattice scale. 

\begin{figure}[htb]
\includegraphics[width = 8.0cm]{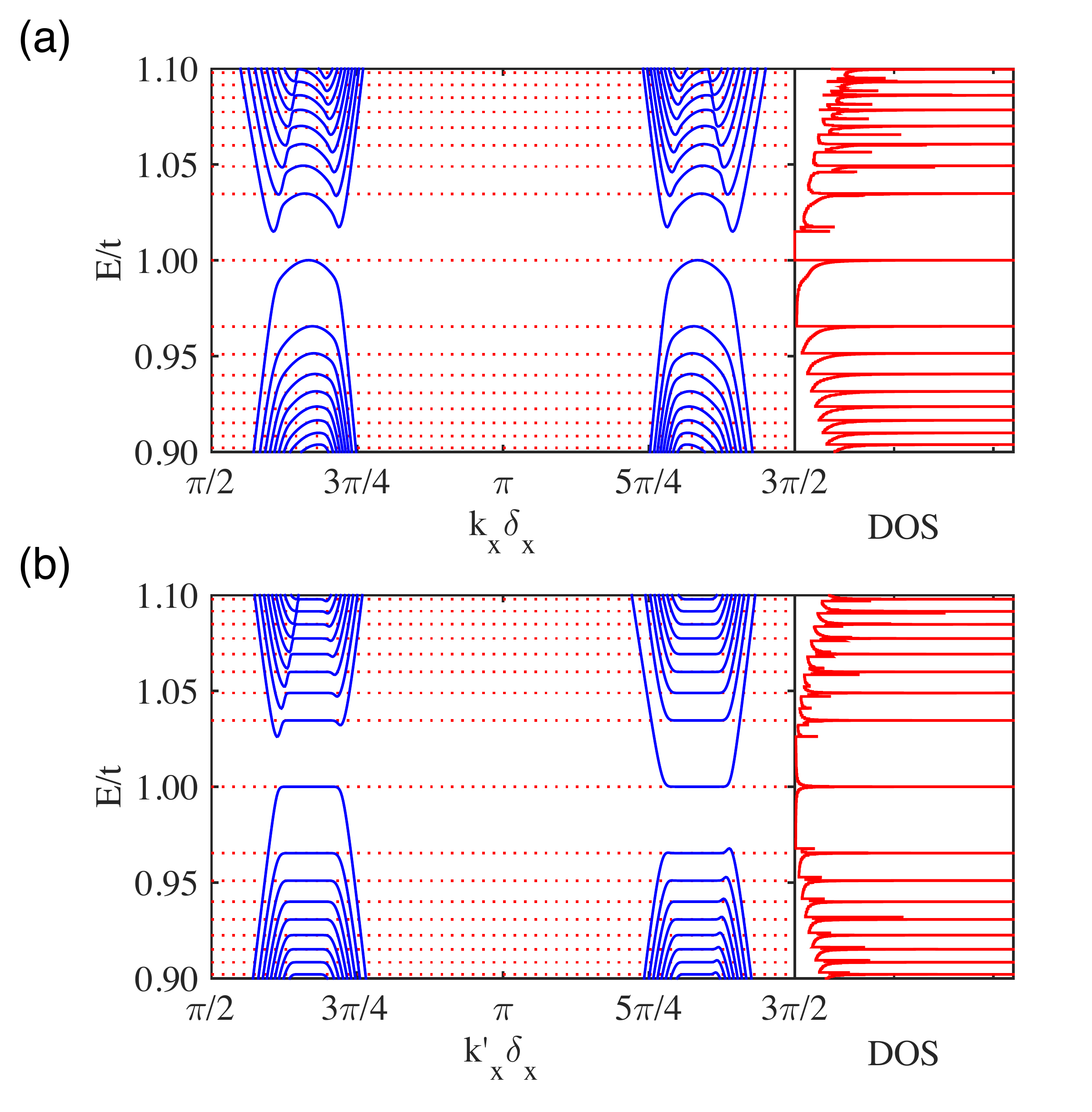}
\caption{Numerically calculated Landau levels of a kagome nanoribbon with a sawtooth edge and a flat edge with $2N=600$ shaded triangles. (a) Landau levels (blue curves) due to a strain-induced pseudomagnetic field (Eq.~\ref{pseudo_B}) with $\sqrt{3}e \mathcal{B} a^2/ 4\hbar =8.66\times 10^{-5}$ flux quanta per shaded triangle. The red dotted lines are the theoretically predicted Landau levels (Eq.~\ref{pseudo_LL}), while the red curve is the numerically calculated DOS. For comparison, the numerically calculated Landau levels and DOS in the presence of an ordinary magnetic field $\bm B = B \hat z$ with $\sqrt{3}e B a^2/ 4\hbar =8.66\times 10^{-5}$ are plotted in (b), where the momentum $k_x'=k_x-\frac{e}{\hbar}B\cdot \sqrt{3}Na$.} \label{Fig3}
\end{figure}

To support our argument, we consider a properly designed lattice deformation characterized by displacement field $\bm u = (\frac{C}{a}xy, -\frac{C}{2a}x^2-\frac{C}{2a}y^2)$ which cancels the strain-induced scalar potential $\mathcal{U}$ but produces a strain-induced vector potential $\mathbfcal{A}=-\eta \hbar C y/\sqrt{3}ea^2 \hat{x}$. The consequent homogeneous pseudomagnetic field is
\begin{equation} \label{pseudo_B}
\mathbfcal{B}=\nabla \times \mathbfcal{A}= \eta \frac{\hbar}{\sqrt{3}ea^2}C \hat z,
\end{equation}
which leads to pseudo Landau levels
\begin{equation} \label{pseudo_LL}
\varepsilon_n=\text{sgn}(n) \sqrt{\bigg|2n\frac{e\mathcal{B}}{\hbar}\hbar v_x^\eta \hbar v_y^{\eta}\bigg|} \qquad n=0, \pm 1, \pm 2, \cdots.
\end{equation}
We have numerically verified these pseudo Landau levels by applying hopping substitution $t_i \rightarrow t_i + \delta t_i$ with $\delta t_i$ listed in Eq.~\ref{delta_t} to the tight-binding Hamiltonian of a nanoribbon with a sawtooth edge and a flat edge. Such a nanoribbon can be obtained by removing the $c_0$ sites in the  nanoribbon illustrated in Fig.~\ref{Fig2}(a); thus contains $2N$ shaded triangles. We emphasize that the profile of the boundary does not affect the pseudo Landau levels, which result from the strain-induced pseudomagnetic field coupled to the Dirac cones deep in the bulk. Indeed, we observe that the pseudo Landau levels (Eq.~\ref{pseudo_LL}) capture the feature of the spectrum and the DOS in the vicinity of Dirac points $\epsilon_{\bm K_\eta, i=1,2}=t$ as illustrated in Fig.~\ref{Fig3}(a). For comparison, we also plot the spectrum and the DOS for the unstrained nanoribbon in the presence of an ordinary magnetic field which coincides with the pseudomagnetic field at valley $K$. Similar results are shown in Fig.~\ref{Fig3}(b).

\begin{figure*}[htb]
\includegraphics[width = 16.0cm]{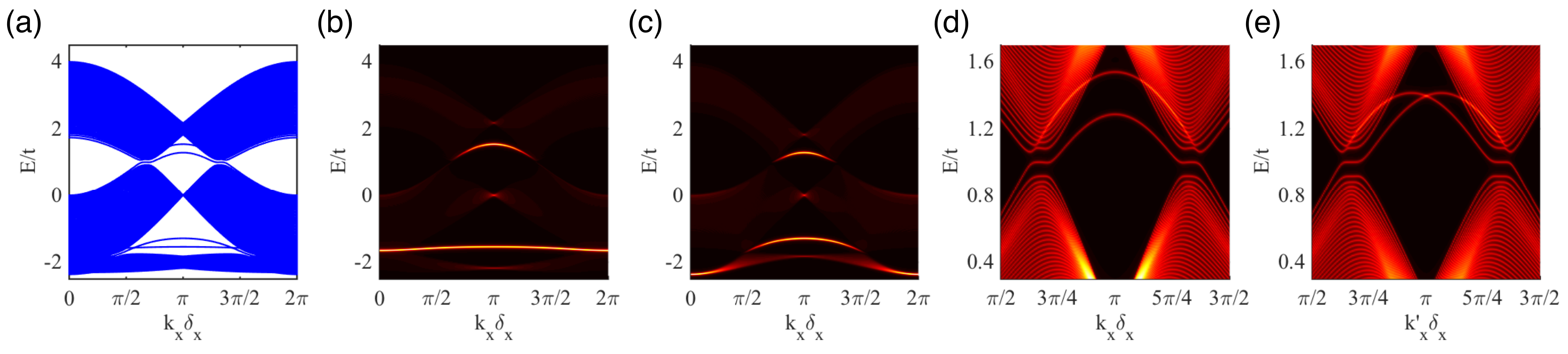}
\caption{Edge state separation in a sawtooth kagome nanoribbon under strain. (a) Band structure of a kagome nanoribbon with a pair of sawtooth edges in the presence of strain. The degenerate edge states in the undeformed nanoribbon are now separated. (b) The spectral function of the upper sawtooth edge. The two edge states are further separated by the strain. (c) The spectral function of the lower sawtooth edge. The spacing of the two edge states is reduced by the strain. (d) Band structure in the vicinity of Dirac points under strain. The Dirac points are broadened by the strain-induced pseudomagnetic field into flat zeroth Landau levels, whose outer (inner) ends are connected by the edge state residing on the upper (lower) sawtooth edge. The edge states are separated in the energy dimension. (e) Band structure in the vicinity of Dirac points under the ordinary magnetic field. The Dirac points extend into flat zeroth Landau levels whose left (right) ends are connected by the edge state hosted by the upper (lower) sawtooth edge. The edge states are thus separated in the momentum dimension. } \label{Fig4}
\end{figure*}

Though the pseudomagnetic field (Eq.~\ref{pseudo_B}) and the pseudo Landau levels (Eq.~\ref{pseudo_LL}) are purely bulk effects, the elastic strain from which they are derived can affect the edge states as well because the displacement field we have employed deforms the two edges differently. Therefore, the energy band degeneracy arising from edge states will generally be lifted. To confirm this, we numerically study a sawtooth kagome nanoribbon, which hosts doubly degenerate edge states connecting the band crossings [Fig.~\ref{Fig2}(c),(d)]. In the presence of strain, we notice that the two edge states residing on the upper (lower) edge of the ribbon become more (less) separated as illustrated in Fig.~\ref{Fig4}(b) [Fig.~\ref{Fig4}(c)]. Specifically, for the edge states between the two dispersive bands, we find that the one on the upper edge always has higher energy than that on the lower edge. This is similar to the edge state separation in bent graphene nanoribbons \cite{stuij2015}. 

To obtain more insights, we take a closer look in the vicinity of Dirac points to see how the edge states emerge from the bulk. Without deformation, the edge states connect Dirac points $K$ and $K'$, similar to the Fermi arcs in Weyl semimetals. When the ribbon is gradually deformed, due to the induced pseudomagnetic field, the two Dirac points at $K$ and $K'$ extend into the zeroth Landau levels, whose outer (inner) ends are connected by the edge state on the upper (lower) sawtooth edge, resulting in edge state separation in the energy dimension as illustrated in Fig.~\ref{Fig4}(d). In the presence of an ordinary magnetic field, the edge state on the upper (lower) edge connects the left (right) ends of the two zeroth Landau levels as illustrated in Fig.~\ref{Fig4}(e), leading to edge state separation in the momentum dimension.

\section{Strain-induced quantum oscillations}
\label{sec4}
In Sec.~\ref{sec3}, we have demonstrated that a properly designed elastic strain can induce a uniform pseudomagnetic field in kagome crystals. The electronic states near the Dirac points are reorganized by the strain into Landau levels, implying that the magnetic transport associated with Landau levels can be reproduced by the applied elastic strain. In the present section, we will numerically study the quantum oscillations in kagome crystals in the presence of the strain-induced pseudomagnetic field. 

Though predicted to be flat as in Eq.~\ref{pseudo_LL}, the actual strain-induced Landau levels in a finite size kagome crystal are dispersive as illustrated in Fig.~\ref{Fig3}(a). This is because the Peierls substitution $\bm q \rightarrow \bm q + \tfrac{e}{\hbar} \mathbfcal{A}$ for the Dirac Hamiltonian under strain (Eq.~\ref{H_strain}) is only rigorously valid at the Dirac points, away from which the higher order correction terms $[O(u_{ij}q_k)]$ in the strain-induced Hamiltonian (Eq.~\ref{dH}) begins to deform the Landau levels. 

To appropriately incorporate the effect from the dispersive Landau levels, we employ the tetrahedron method \cite{blochl1994} to calculate the quantum oscillations of the DOS and the electric conductivity. In particular, we first divide the 1D Brillouin zone into a set of discretized momenta $k_x^i$ with $i=1,2,\cdots,i_{\text{max}}$ such that each interval $[k_x^i, k_x^{i+1}]$ is sufficiently narrow. Therefore, the Landau levels in such an interval can be approximated to disperse linearly
\begin{equation}
\varepsilon_n(k_x) = \frac{\varepsilon_n^{i+1}-\varepsilon_n^i}{k_x^{i+1}-k_x^i} k_x + \frac{k_x^{i+1}\varepsilon_n^i - k_x^i \varepsilon_n^{i+1}}{k_x^{i+1}-k_x^i},
\end{equation}
where we have denoted $\varepsilon_n^i=\varepsilon_n(k_x^i)$ for transparency. Consequently, the DOS is 
\begin{multline} \label{dos}
g(\mu)= \frac{L_x}{2\pi} \sum_n \sum_{i=1}^{i_{\text{max}}-1} \int_{k_x^i}^{k_x^{i+1}} dk_x \delta[\mu-\varepsilon_n(k_x)]
\\
= \frac{L_x}{2\pi} \sum_n \sum_{i=1}^{i_{\text{max}}-1} \frac{k_x^{i+1}-k_x^i}{\varepsilon_n^{i+1}-\varepsilon_n^i} [\Theta(\mu-\varepsilon_n^i)-\Theta(\mu-\varepsilon_n^{i+1})],
\end{multline}
where $L_x$ is the length of the ribbon and $\Theta$ is the Heaviside step function. And the electric conductivity can be calculated using the semiclassical method \cite{ashcroft1976} as
\begin{multline} \label{cond}
\sigma_{xx}(\mu) 
\\
=\frac{L_x}{2\pi} \sum_n \sum_i^{i_{\text{max}}-1} \int_{k_x^i}^{k_x^{i+1}} dk_x e^2 \tau_n[\varepsilon_n(k_x)] [v_n^x(k_x)]^2 \bigg( -\frac{\partial f}{\partial \varepsilon} \bigg)_{\varepsilon_n(k_x)}
\\
\overset{T\rightarrow 0}{=} \frac{e^2L_x}{2\pi\hbar^2} \tau(\mu) \sum_n \sum_i^{i_{\text{max}}-1} \frac{\varepsilon_n^{i+1}-\varepsilon_n^i}{k_x^{i+1}-k_x^i} [\Theta(\mu-\varepsilon_n^i)-\Theta(\mu-\varepsilon_n^{i+1})],
\end{multline}
where $v_n^x(k_x) = \frac{1}{\hbar} \frac{\partial \varepsilon_n(k_x)}{\partial k_x}$ is the band group  velocity and $f(\varepsilon)=[e^{(\varepsilon-\mu)/k_BT}+1]^{-1}$ is the Fermi distribution function. We have assumed temperature $T=0$ and identical relaxation time $\tau_n(\mu)=\tau(\mu)$ for all the Landau levels. Explicitly, the scattering rate can be approximated by the lowest order Born approximation \cite{doniach1998}
\begin{equation} \label{tau}
\frac{1}{\tau(\mu)}=2\pi g(\mu) n_{\text{imp}} C,
\end{equation}
where $n_{\text{imp}}$ is the impurity concentration and $C$ depends on the strength of the scattering.

\begin{figure}[htb]
\includegraphics[width = 8.0cm]{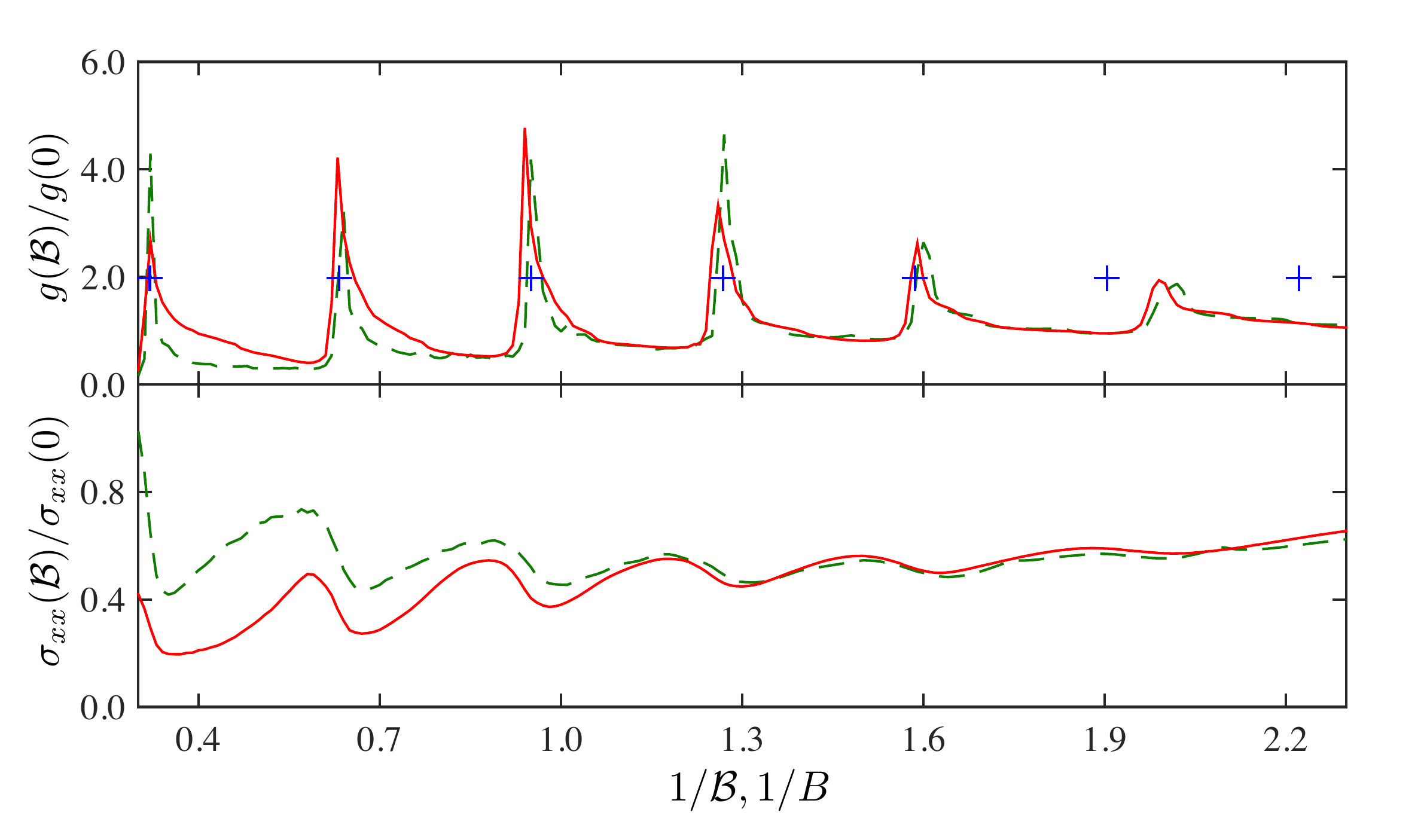}
\caption{Quantum oscillations of kagome crystals. The upper panel shows the quantum oscillations of the DOS at the Fermi level $\mu=0.9565t$ resulting from the strain-induced pseudomagnetic field (red solid curve) and the ordinary magnetic field (green dashed curve). The blue crosses mark the value of $1/\mathcal{B}$ and $1/B$ where Landau levels hit the Fermi level. The lower panel shows the quantum oscillations of the electric conductivity, i.e., Shubnikov-de Haas (SdH) oscillations, caused by the strain-induced pseudomagnetic field (red solid curve) and the ordinary magnetic field (green dashed curve). Both conductivity curves are broadened by convolving a Lorentzian of width $\delta=0.001t$ to consider the impurity effect in the relaxation time (Eq.~\ref{tau}). The DOS curves are also broadened but by a much smaller Lorentzian width $\delta'=0.08\delta$ in order to better characterize the (pseudo) magnetic fields at which Landau levels coincide with the Fermi level. The (pseudo) magnetic field is measured in the unit of $10^{-4}\hbar/ea^2$.} \label{Fig5}
\end{figure}

We have numerically calculated the DOS and the electric conductivity at different pseudomagnetic fields and find they both exhibit oscillations periodic in $1/\mathcal{B}$ at the Fermi level $\mu=0.9565t$ as illustrated in Fig.~\ref{Fig5}. This is because the Landau levels successively pass through the Fermi level when their spacing is continuously tuned by the varying pseudomagnetic field, resulting in periodic population of electrons on the Fermi level. We have also verified other Fermi energies and observe similar oscillations. For comparison, the quantum oscillations of the DOS and the electric conductivity due to the ordinary magnetic field are overlaid in Fig.~\ref{Fig5}. The match of the periodicity further confirms the similarity of the strain-induced pseudomagnetic field to the ordinary magnetic field.

\section{Conclusions}
\label{sec5}
In this paper, by a proper projection of the sublattice basis, we have studied the Dirac physics of kagome crystals focusing on its response to the applied elastic strain and the associated magnetic transport in the form of quantum oscillations. We first analyze the band structure of a kagome crystal with only nearest neighbor terms. By projecting the sublattice basis Hamiltonian onto the space spanned by the eigenvectors associated with the Dirac points, we are able to drop off the degree of freedom related to the flat band and analytically extract the Dirac physics, which is also numerically simulated for nanoribbons with sawtooth edges and zigzag edges, respectively. For both cases, we find two pairs of doubly degenerate edge states emerging from and terminating at band crossings, i.e., Dirac points between the two dispersive bands and the quadratic band touchings between the flat band the lower dispersive band.

Then, we incorporate elastic strain by hopping integral substitution and elucidate that the most important effect of the strain to the bulk states is to shift the Dirac points oppositely. Therefore, a properly designed strain can be interpreted as a uniform chiral pseudomagnetic field resulting in Landau quantization. By numerically studying a sawtooth kagome nanoribbon, we find the strain modulates the edges differently thus lifting the edge state degeneracy. In particular, the edge states connecting the Dirac points  become completely separated in the energy dimension due to the unique way they connect to the zeroth Landau levels broadened from the Dirac points. Similar separation, however, is absent in the presence of the ordinary magnetic field due to the different connection of the edge states to the zeroth Landau levels, reflecting the fundamental difference between the strain-induced chiral pseudomagnetic field and the ordinary magnetic field.

Lastly, we study the DOS and the longitudinal electric conductivity of a kagome nanoribbon under different values of the strain-induced pseudomagnetic field. The tetrahedron method is adopted in order to appropriately treat the Landau level dispersion resulting from the higher order correction [$O(u_{ij}q_k)$] to the strain-induced Hamiltonian. Though departing from the ideal flat Landau levels, the strain-induced dispersive pseudo Landau levels can produce quantum oscillations in both the DOS and the longitudinal electric conductivity.

To experimentally implement the strain-induced quantum oscillations, we first require a kagome crystal such as Co$_3$Sn$_2$S$_2$ \cite{yin2019, liu2018, wang2018}, FeSn \cite{kang2019}, and Fe$_2$Sn$_3$ \cite{ye2018, ye2019, yin2018} in which ordinary magnetic field Shubnikov-de Haas oscillations and de Haas-van Alphen oscillations have been observed \cite{liu2018, kang2019, ye2019}. Secondly, the candidate materials must be sufficiently flexible to sustain the strain designed for Landau levels. However, the mechanical properties regarding the flexibility of the aforementioned materials are not complete and further experimental work is needed to verify whether one or more of these materials are suitable for the strain-induced quantum oscillation experiment. Lastly, in order to have a scanning pseudomagnetic field, the strain should be tuned continuously with ease. Unfortunately, the strain we have engineered is not of this type. Nevertheless, it is worth noting that the $x$ component of the displacement field $u_x=\frac{C}{a}xy$ originates from a circular bend, which can be tuned easily \cite{liu2017a}, while the existence of the $y$ component $u_y=-\frac{C}{2a}(x^2+y^2)$ aims at canceling the strain-induced pseudoelectric field and preserving the translational symmetry. Moreover, without $u_y$, $u_x$ itself can also generates a uniform pseudomagnetic field $\mathbfcal{B}= \eta \frac{\hbar C}{\sqrt{3}ea^2}\hat z$. We thus argue that a circular bend lattice deformation should be sufficient for the experiment of strain-induced quantum oscillations.

\begin{acknowledgments}
The authors are indebted to S. Fujimoto and Z. Shi for insightful discussions. We particularly thank R. Moessner and P. A. McClarty for the valuable suggestions for this work.
\end{acknowledgments}

\bibliography{Kagome}
\clearpage
\end{document}